\documentclass{elsart}
\journal{New Astronomy}

\usepackage{epsfig}

\usepackage{amssymb}

\def\n{{\bf n}}

\def\bi{\begin{itemize}}
\def\ei{\end{itemize}}

\def\n2{N$_{2}$}
\def\4he{$^{4}$He}
\def\cm3{cm$^3$}

\newcommand{\arcdeg}{^\circ}

\begin{document}

\begin{frontmatter}



\title{CMB Optical Depth Measurements: Past, Present, Future}


\author{Brian Keating$^1$ and Nathan Miller$^1$}
\address{$^1$Department of Physics, University of California, San Diego, Mail Code 0424, La Jolla, CA,
92093-0424, United States of America\\
E-mail: bkeating@ucsd.edu, nmiller@physics.ucsd.edu}

\begin{abstract}
The polarization of the cosmic microwave background (CMB) is
encoded with exactly the same cosmic information as the CMB's
temperature anisotropy. However, polarization has the additional
promise of accurately probing the reionization history of the
universe and potentially constraining, or detecting, the
primordial background of gravitational waves produced by
inflation. We demonstrate that these two CMB polarization goals
are mutually compatible. A polarimeter optimized to detect the
inflationary gravitational wave background signature in the
polarization of the CMB is well situated to detect the signatures
of realistic first-light scenarios. We also discuss current
results and prospects for future CMB polarization experiments.
\end{abstract}

\end{frontmatter}

\section{Introduction}

The CMB possesses a wealth of cosmological information owing to
its origin 400,000 years after the Big Bang -- an epoch when
linear cosmological perturbation theory holds. However, the CMB
also is imprinted by ``first-light" sources at later times that
reveal the dynamics of reionization. Even detecting the transition
from partial to full-reionization in (physically more plausible)
non-instantaneous reionization scenarios is conceivable. The
first-year WMAP results demonstrate a nearly $5\sigma$ detection
of a reionization feature in the temperature-polarization
cross-correlation power spectrum. This detection, along with
WMAP's detection of the Thomson optical depth, was the first
demonstration of CMB polarization's ability to go ``beyond
parameter estimation" -- making precision measurements of a
cosmological epoch which was previously shrouded in mystery.
WMAP's detection was also a tantalizing hint at CMB polarization's
promise to probe vitally important cosmological parameters that
are essentially unobservable using temperature anisotropy alone.

However, as with many groundbreaking experiments, WMAP's results
have generated even more intriguing questions.  In particular, why
is the Thomson optical depth so much higher for the CMB than that
implied by the QSO Gunn-Peterson measurements, e.g.
\cite{becker2001,djorgovski2001,fan2002}? Ultimately, the optical
depth due to first light sources measured by CMB polarimeters must
be reconciled with the Gunn-Peterson absorption feature
measurements in the spectra of distant quasars. Several attempts
at this reconciliation have been proposed, some of which feature
multiple reionizations \cite{cen2003}, or a period of partial
ionization \cite{kaplinghat2003}. Interestingly, it appears that
these reionization scenarios, which translate to structure
formation scenarios, can indeed be constrained observationally by
CMB polarization.

In this paper we discuss the prospects of upcoming ground,
balloon, and space CMB polarimeters to measure both the Thomson
optical depth and constrain new features in the polarization power
spectra characteristic of realistic reionization scenarios. The
implications of intrinsic systematic effects such as foregrounds
and cosmic variance are also discussed.

\section{CMB Polarization}

Much of the wealth of cosmological information encoded in CMB
polarization is only observable at medium to large angular scales
($>0.5\arcdeg$). Essentially \textit{all} information on the
classical cosmological parameters of the $\Lambda$CDM model,
\textit{e.g.}, the baryon density, $\Omega_b$, the Hubble
parameter, $h$, the scalar power spectrum's power law index,
$n_s$, the amplitude of the primordial power spectrum, $A$, the
tensor-to-scalar ratio, $r$, etc, can be obtained from CMB
polarization at angular scales
 $ >0.5\arcdeg$. Because the temperature anisotropy is much larger than the polarization signal it \emph{can}
produce higher confidence measurements of the classical
cosmological parameters than polarization. However, there are
several parameters which \emph{only} polarization can reveal. In
particular, only CMB polarization can probe the imprint of the
inflationary gravitational wave background (GWB) on the CMB. This
polarization signature occurs at angular scales of $\sim2\arcdeg$.
Reionization primarily affects CMB polarization on angular scales
$>4\arcdeg$. We note that the grad-mode polarization, $C_\ell^E$,
peaks at $\ell \simeq 1000$ corresponding to $\simeq 10^\prime$
scales, and that gravitational lensing of the CMB grad-mode
polarization by large scale structure produces non-negligible
curl-mode polarization $C_\ell^B$ at the same $\simeq 10^\prime$
scales \cite{hu2002}.

\section{Current Results}
\begin{figure}[htb]
\centerline{\psfig{file=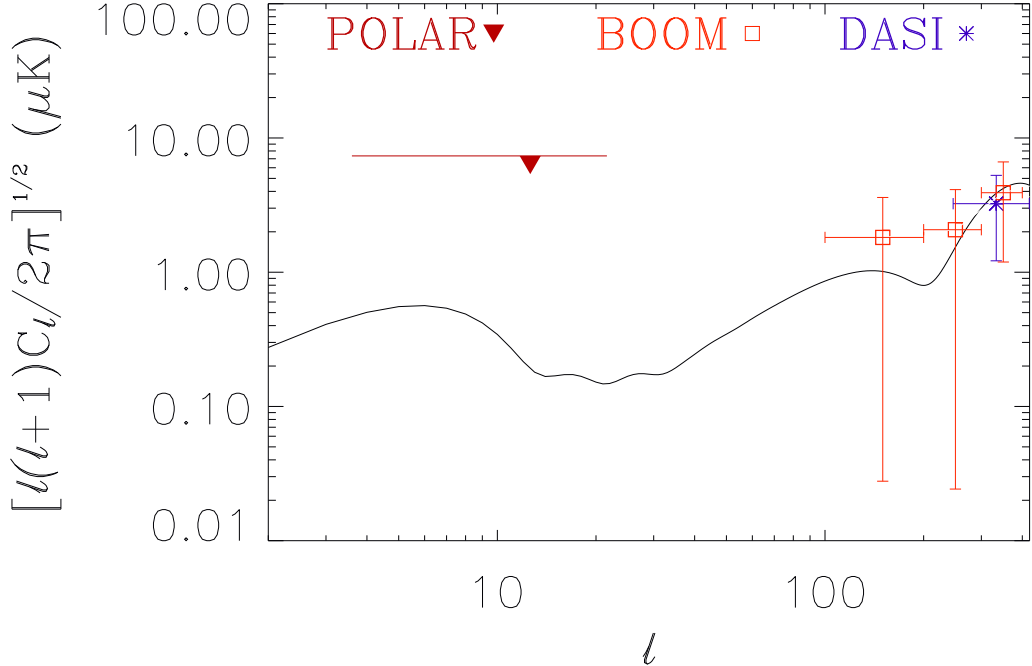, width=5.9in, angle=0}}
\caption{Measurements of the large-angular scale gradient-mode
polarization power spectrum $C_{\ell}^{E}$ for $0.5\arcdeg
\Leftrightarrow \ell < 400$. The solid line is the polarization
power spectrum for the WMAP best fit cosmological model, with
$\tau=0.17$ \cite{spergel2003}.} \label{fig:emode_limits}
\end{figure}

\begin{figure}[htb]
\centerline{\psfig{file=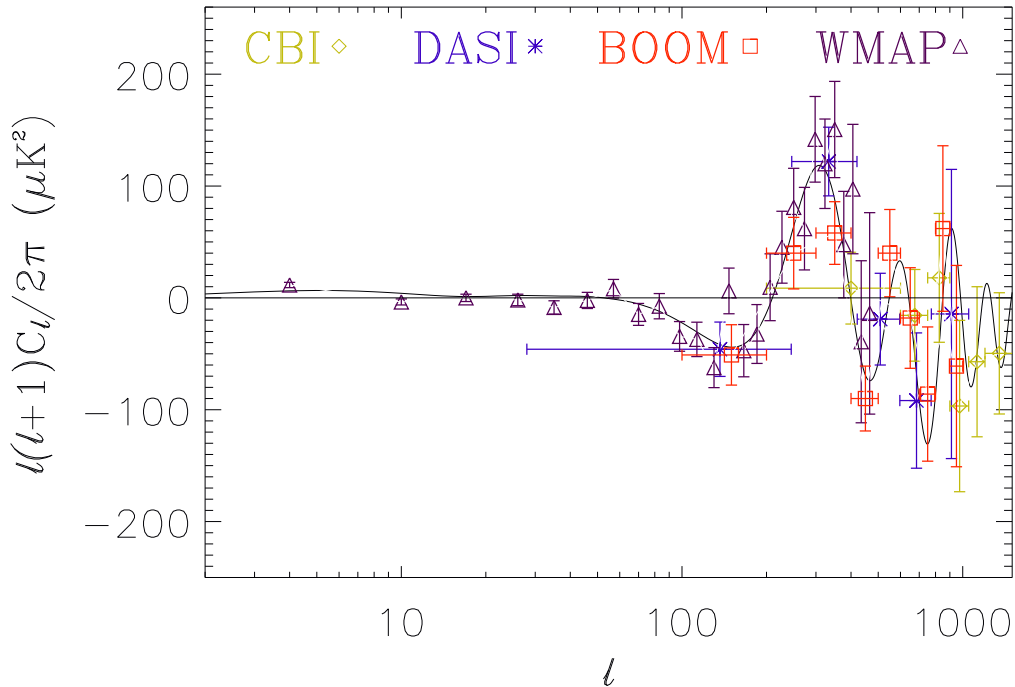, width=5.9in, angle=0}}
\caption{Measurements of the temperature-polarization
cross-correlation power spectrum $C_{\ell}^{TE}$. The solid line
is the power spectrum for the WMAP best fit cosmological model,
with $\tau=0.17$  \cite{spergel2003}.} \label{fig:temode_limits}
\end{figure}

Figures \ref{fig:emode_limits} and \ref{fig:temode_limits} show
the current detections of the grad-mode polarization and the
polarization-temperature cross correlation. Currently, only
DASI\cite{Leitch2005} and BOOMERANG\cite{Montroy05} have detected
grad-mode polarization at angular scales
$<0.5\arcdeg\Leftrightarrow\ell < 400 $, and no experiments have
detected the grad-mode polarization for $\ell < 100$. The only
other experiments that have detected the grad-mode polarization
are CBI\cite{Readhead04} and CAPMAP\cite{barkats2005}, though at
smaller angular scales (higher $\ell$) than are considered in this
paper.

The primary effects of reionization are encoded in the grad-mode
and temperature polarization cross-correlation at $\ell \lesssim
50$. In the absence of detections, the most stringent constraint
in this range of multipoles is currently $C_\ell^E < 8$ $\mu K$ at
95\% confidence reported by POLAR\cite{Keat01} for $2 < \ell < 20$
(assuming no B-modes). There are many more detections of the
temperature-polarization cross correlation than of the grad-mode
polarization as figure \ref{fig:temode_limits} demonstrates.
WMAP\cite{kogut2003} reports a large number of highly significant
detections, especially at low-$\ell$ due to WMAP's ability to map
most of the sky. CBI, DASI, and BOOMERANG\cite{Piacentini2005}
have also detected the cross-correlation spectrum, mainly at much
smaller angular scales than WMAP. A complete description of
reionization will require significant detections of CMB E-mode
polarization for $\ell<50$. An ancillary benefit of reionization
is that it boosts the primary curl-mode power spectrum
significantly near $\ell = 10$. Due to reionization, a more
stringent limit on the tensor-to-scalar ratio, $r$, in the
presence of lensing can be obtained than that calculated in
\cite{knox2002,kesden2002}.

\section{The effect of reionization}

\begin{figure}[t]
\includegraphics[height=7.5cm]{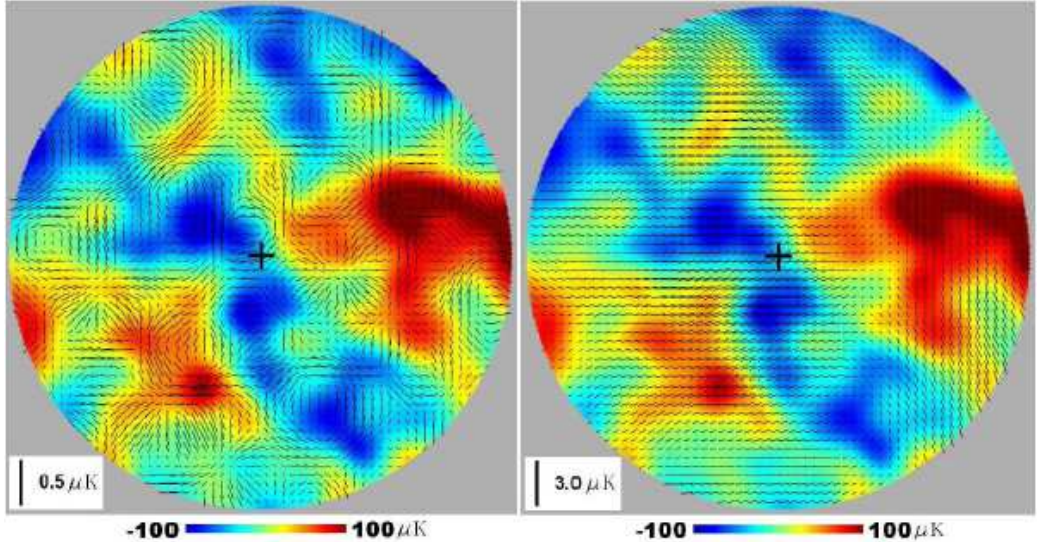}
\caption{\label{fig:polmaps} Temperature (color scale) and
polarization (lines) realizations in real-space smoothed with a
$4\arcdeg$ beam. The left panel shows a simulation with no
reionization and the right panel shows the effects of
instantaneous reionization with $\tau=0.17$. Reionization does not
noticeably affect the large scale temperature pattern but it
produces roughly ten-times larger polarization at large scales.
These figures have no curl-mode power. The maximum gradient-mode
polarization in the right panel is $\simeq 800$ nK. Figures
courtesy of Eric Hivon.}
\end{figure}

Reionization produces free-electrons which Thomson-scatter CMB
photons and produce CMB polarization. The scattering damps the CMB
temperature in direction $\hat{\textbf{n}}$ as
$T^\prime(\hat{\textbf{n}})=e^{-\tau}T(\hat{\textbf{n}}),$ where
$\tau$ is the Thomson optical depth. Reionization, therefore,
damps the anisotropy power spectrum as
$C_\ell^{T^\prime}=e^{-2\tau}C_\ell^T$. The effect of $\tau$ on
the temperature anisotropy power spectrum is completely degenerate
with the primordial power spectrum's amplitude, $A$. Fortunately,
a new feature, e.g. \cite{zaldarriaga1997,keating1998}, in the
polarization power spectra occurs at large angular scales which
breaks the degeneracy between normalization and optical depth. The
degeneracy (for the CMB temperature anisotropy) and its breaking
(for CMB polarization) is nicely illustrated in (real-space) in
figure \ref{fig:polmaps}.

\section{A More Realistic Cosmic Variance Limit.}

An all-sky experiment must confront galactic contamination, which
can only be subtracted  to finite precision. For example, the WMAP
team used multiple frequency channels in combination with a
sky-cut that left $85\%$ of the sky remaining \cite{kogut2003}.
This cut also introduced correlations between multipoles at the
few percent-level, and a total anti-correlation of 12.4\%. Even
with sophisticated foreground modelling and multiple frequency
coverage the challenges of achieving polarization fidelity at the
100 nK level over $>10\arcdeg$ angular scales -- required, for
example, to detect the signatures of non-instantaneous
reionization scenarios (e.g, \cite{kaplinghat2003,holder2003}) --
are significant.

As a simple, but admittedly conservative, foreground mitigation
technique we consider a full-sky map and and omit galactic
latitudes within $|b|<b_{cut}$ of the galactic plane. This leaves
two circular ``cap" regions remaining. This strategy affects the
recovery of reionization information in two ways. First it reduces
the resolution in $\ell$-space compared to a true full-sky
experiment, leaving a residual correlation between multipoles.
Secondly, it increases cosmic variance on the measurement of
$C_\ell^E$, since the amount of observed sky, $f_{sky}$ decreases:
$\delta C_\ell \propto \frac{1}{\sqrt{f_{sky}}}$.

The effect of this simple foreground mitigation strategy is
apparent in parameter estimation. If one (conservatively) insists
on using uncorrelated multipoles to estimate parameters, the
integrated signal-to-noise ratio $\displaystyle\sum_{\ell=2, 4,
6\ldots}^{\ell_{cutoff}} \frac{C_\ell^E}{\delta C_\ell^{E}}$ is
reduced significantly. Here $\delta C_\ell^{E}$ is the cosmic
variance limited uncertainty on $C_\ell^{E}$, including the
inflation of the cosmic variance due to the cut itself, and
$\ell_{cutoff}$ is the highest multipole afforded by the angular
resolution of the telescope or other filtering.

As a numerical example, we consider a galaxy-cut which excludes
data with $b_{cut}=20\arcdeg$, corresponding to $f_{sky}=75\%$ --
not far off from WMAP's $f_{sky}$. However, since we insist that
the total anti-correlation between multipoles is $\ll1\%$, the
$\ell$-space resolution degrades from $\delta \ell = 1$ to $\delta
\ell = 2$. The quadrupole, and all even multipoles are completely
uncorrelated.

We find that the signal-to-noise is more than halved relative to a
true all-sky experiment $\displaystyle\sum_{\ell=2, 3,
4\ldots}^{\ell_{cutoff}} \frac{C_\ell^E}{\delta C_\ell^{E}}$ . The
halved signal-to-noise ratio of the power spectra does not
\textit{necessarily} translate into a doubling of parameter
errors, however. A Fisher matrix analysis, \textit{e.g.}
\cite{Dodelson2003}, must be performed to assess the degradation
of parameter sensitivity.

\begin{figure}[htb]
\centerline{\psfig{file=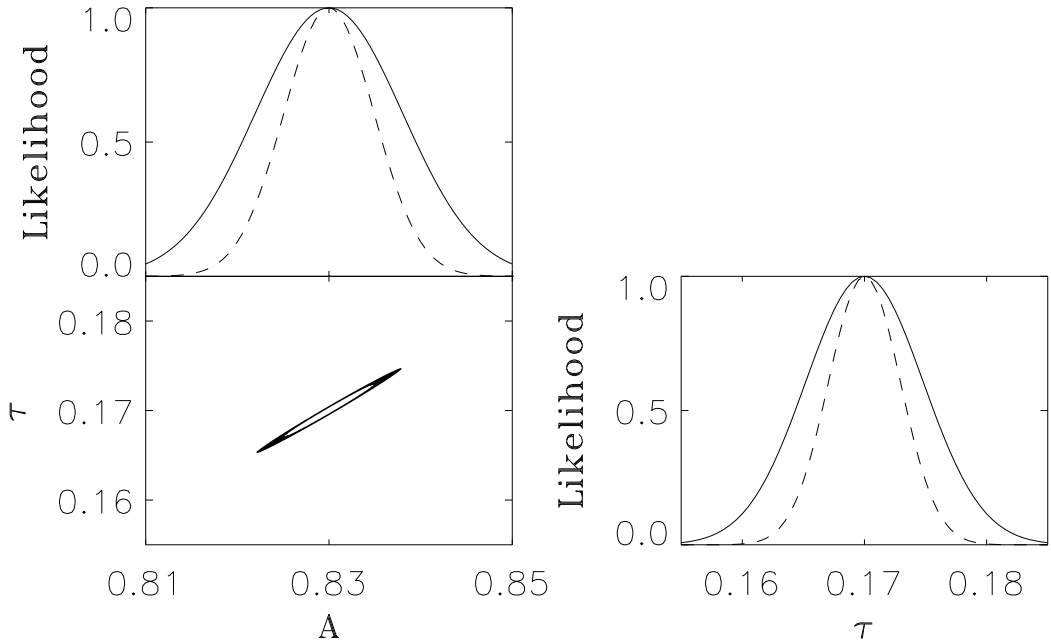, width=5.9in, angle=0}}
\caption{Individual and joint likelihood functions for the cut-sky
and full-sky observations discussed in the text. The errors on
$\tau$ and $A$ from the 1D likelihoods for the $b_{cut}=20\arcdeg$
simulation (solid) are inflated by $\sim 65\%$ over the true
full-sky experiment (dashed) due to increased cosmic variance and
decreased multipole space resolution. The 2D joint likelihood
contours show the full-sky survey's error volume (white) inside
the the cut-sky data's region (black). The contours indicate where
the likelihood falls to $1/\sqrt{e}$ of the maximum likelihood for
each experiment.} \label{fig:tauas}
\end{figure}

A parameter estimation indicates that the new cosmic variance
limit on $\tau$ from the $b_{cut}=20\arcdeg$ cut-sky data is
$\sigma_\tau=0.0048$. This is compared to our Fisher matrix
results for the ideal full-sky cosmic variance limit:
$\sigma_\tau=0.0029$ -- an increase of 65\%. This latter limit
agrees nearly exactly with the results of Holder et al.
(2003)\cite{holder2003}. Due to the nearly degenerate correlation
between $\tau$ and the amplitude of the primordial power spectrum,
the increase in $\sigma_\tau$ increases the uncertainty on $A$ as
well. Figure \ref{fig:tauas} shows the individual and joint
likelihoods for $\tau$ and $A$ for the full-sky and
$b_{cut}=20\arcdeg$ simulations.

Additionally, more realistic cut-sky observing strategies have
lower $\ell$-space resolution than all-sky surveys which may
preclude the detection of features in the polarization power
spectra that characterize the dynamics of reionization
\cite{kaplinghat2003,holder2003}. This may ultimately limit the
precision with which the dynamics of non-instantaneous
reionization can be reconstructed. Furthermore Kaplinghat et
al.\cite{kaplinghat2003} estimate that the sensitivity of a CMB
polarimeter to the redshift of the onset of reionization is
$\sigma_{z_{ri}}=333\sigma_\tau/\sqrt{z_{ri}}$. This implies that
$\sigma_{z_{ri}}$ for a more realistic observation strategy (i.e.,
cut-sky) will be determined $~65\%$ less precisely than an ideal
truly all-sky measurement. These effects will thus be more
pronounced as the sensitivity of CMB polarimeters increase.

This more realistic cosmic variance limit may also strengthen the
motivation for ground-based CMB polarization probes of
reionization which, due to their significantly reduced sky
coverage relative to a satellite or balloon, have received
comparatively little attention.

\section{A Ground Based Reionization Probe}

In this section we explore how well a ground based polarimeter can
probe reionization. We seek to quantify the observational
requirements to detect $\tau$, as well as to constrain realistic
reionization models consistent with both WMAP and quasar
absorption measurements. In particular we assess the requirements
to detect a two-step reionization model suggested by
\cite{kaplinghat2003}.

Generally speaking, an instrument optimized to detect
gravitational waves will significantly constrain the detailed
reionization history, whereas the converse is not necessarily
true. For fixed system sensitivity this has implications for both
the optical design and observation strategy of the experiment.

Here we will consider a medium angular-scale, ground-based
polarimeter's ability to detect $\tau$ and constrain
non-instantaneous reionization scenarios. As a toy experiment, we
consider a polarimeter with $1\arcdeg$ resolution and system
sensitivity
($\textsf{NET}_{\textsf{syst}}=\textsf{NET}_{\textsf{pixel}}/\sqrt{N_{\textsf{pixel}}}
= 70 \mu K s^{1/2}$, observing for one year. The system
sensitivity for polarization is $\sqrt{2}$ times higher
$\textsf{NEQ}=\sqrt{2}\textsf{NET}$. These detector requirements
are well within reach of the current generation of CMB
polarimeters.

The problem of optimizing of CMB polarimeters for detection of the
inflationary GWB was first studied by \cite{jaffe2000}, and
polarimeter optimization in the presence of the gravitational
lensing foreground has been studied by Lewis, Challinor, and Turok
\cite{lewis2002}. Lewis \textit{et al.} find there is a broad
maximum sensitivity achieved by mapping a circular cap of radius
$\Theta=20\arcdeg$, which corresponds to an $\ell$-space
resolution of $\delta\ell \simeq 5$. In the absence of galactic
foregrounds, our toy experiment is capable of detecting the curl
mode signal imprinted by primordial gravitational waves with a
tensor-to-scalar ratio $r=0.12$ at 95\% confidence, with no
priors. This is roughly \emph{\textit{ten-times lower}} than the
limit on $r$ from WMAP, also with no priors \cite{spergel2003}.
This detection (or limit) would permit discrimination between a
broad class of inflationary models
\cite{kinney1998,LiddleLyth2000,kinney2003}. An exciting ancillary
benefit of this experiment is that it is also sensitive to
reionization.
\begin{figure}[htb]
\begin{center}
\begin{tabular}{c}
\includegraphics[height=7cm]{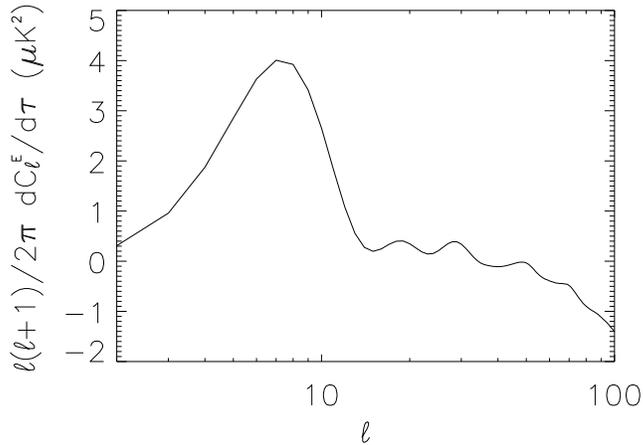}
\end{tabular}
\end{center}
\caption{\label{fig:derive_e} The susceptibility of the E-mode
power spectrum to changes in $\tau$. The derivatives were computed
using a second-order finite difference with steps in $\tau$ of
$\delta \tau=0.001$, around a fiducial value of $\tau=0.17$,
consistent with WMAP. The dominant effect of $\tau$ is at
$\ell<10$, but a non-trivial amount remains for $10<\ell<40$ where
cosmic variance effects can be dramatically reduced, even for a
ground based experiment.}
\end{figure}

\begin{figure}[htb]
\begin{center}
\begin{tabular}{c}
\includegraphics[height=7cm]{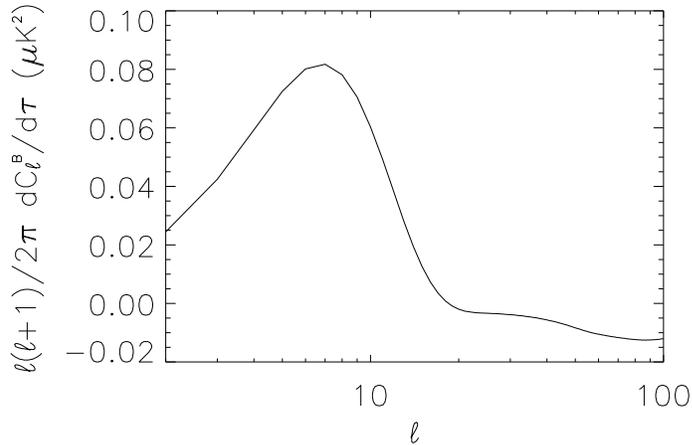}
\end{tabular}
\end{center}
\caption{\label{fig:derive_bb}Same as Fig. \ref{fig:derive_e}, but
for curl-mode polarization.}
\end{figure}

To assess the power of the toy experiment to constrain the Thomson
optical depth we consider the susceptibility of the E-mode power
spectrum to changes in $\tau$: $\partial C^E_\ell/\partial\tau$.
From the grad-mode (Fig. \ref{fig:derive_e}) and curl-mode
derivatives (Fig. \ref{fig:derive_bb}) we see that essentially all
the information on $\tau$ comes from $\ell<50$. If we note that
the peak in the curl-mode power spectrum occurs at $\ell \simeq
90$, we see that the scale for completely resolving the effects of
$\tau$ is only two-times larger than that required to resolve the
curl-mode structure. While most of the $\tau$-constraint comes
from $\ell<10$, a non-negligible amount comes from $10<\ell<40$.
There is significant power beyond the peak of the low-$\ell$ bump.
A Fisher matrix analysis for the toy model indicates its
sensitivity to $\tau$ is $\sigma_\tau=0.08$, competitive with the
one-year WMAP results \cite{spergel2003} with no priors.

\begin{figure}[htb]
\begin{center}
\begin{tabular}{c}
\includegraphics[height=8.25cm]{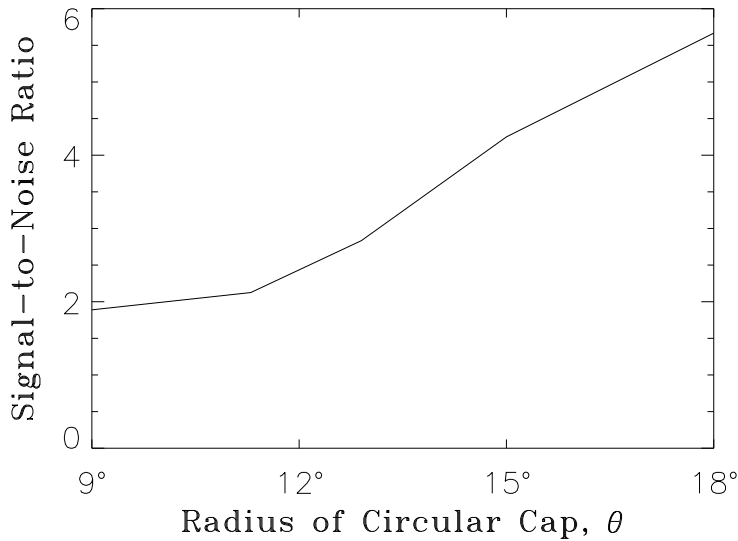}
\end{tabular}
\end{center}
\caption{\label{fig:tausens}Signal-to-noise ratio for a ground
based polarimeter to detect reionization, assuming $\tau=0.17$.
The plot shows the effect of increasing the angular radius of the
circular survey, $\Theta$, with constant system sensitivity
$\textsf{NET}_{syst}=70 \mu$K$s^{1/2}$. All strategies except
$\Theta=9\arcdeg$ exclude the first multipole bin from the
parameter estimation. Lewis, Challinor, and Turok\cite{lewis2002}
show that $\Theta=18\arcdeg$ is close to optimal sky coverage to
detect the inflationary GWB using CMB curl-mode polarization.}
\end{figure}

It is instructive to explore how the sensitivity to $\tau$ changes
as a function of sky coverage. In table 1 and figure
\ref{fig:tausens} we demonstrate how $\tau-$sensitivity,
$\sigma_\tau$, depends on $\ell$-space resolution. The
$\ell$-space resolution scales as $\delta\ell \simeq
180\arcdeg/2\Theta $. With one exception, for all of the
observation strategies we consider we exclude the first $\ell$-bin
which is, in principle, observable. This results in $1^{st}
\ell$-bin$ =2\delta\ell$. This conservatism is frequently
justified since the first $\ell$-bin often is contaminated by an
imperfect foreground template and/or by subtraction of the
DC-level of the map. WMAP's purportedly ``anomalously-low" first
$\ell$-bin, $\ell=2$ has received much attention, perhaps
prematurely. However, some have speculated on the potential for
foreground contamination \cite{schwarz2004} of, or even intriguing
new physics\cite{dore2004} in, the quadrupole's low value.

\begin{table}[t]
\centerline{
\begin{tabular}{|ccccc|}
    \hline
$\Theta$ & $1^{st} \ell$-bin &  $\delta\ell$    &   $f_{sky}$  & $\sigma_\tau$  \\
    \hline
$18\arcdeg $    & 10   & 5 & 2.4\%  & 0.03 \\
$15\arcdeg $   & 12   & 6 & 2.1\%  & 0.04 \\
$12.9\arcdeg$  & 14    & 7 & 1.6\%  & 0.06 \\
$11.3\arcdeg $ & 16    & 8 & 1.3\%  & 0.08 \\
$9\arcdeg   $& 10   & 10 & 0.8\%  & 0.09 \\
    \hline
\end{tabular}}
\caption{Sensitivity to optical depth, $\tau$, versus observation
strategy, parameterized by radius of circular cap, $\Theta$,
observed by the toy experiment described with system
$\textsf{NET}_{syst}=70 \mu$K$s^{1/2}$. For all of the observation
strategies, except $\Theta=9\arcdeg$, the first $\ell$-bin is
$2\delta\ell$.\label{tablel}}
\end{table}

\begin{figure}[h]
\centerline{\psfig{file=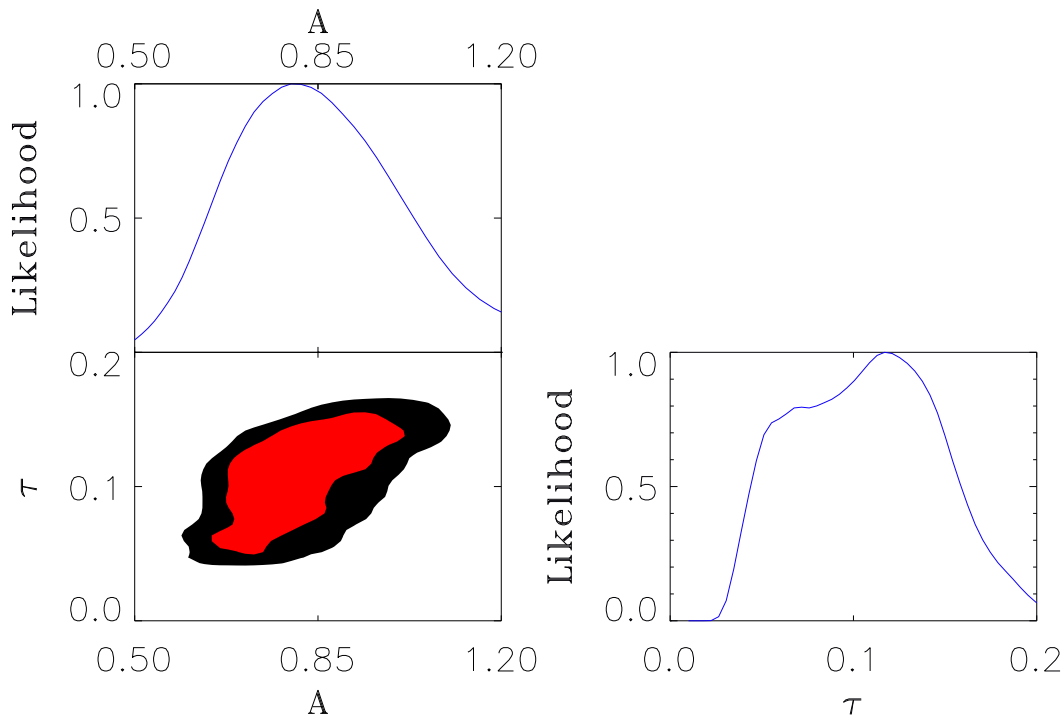, width=5.9in, angle=0}}
\caption{Individual and joint likelihood functions for the toy
experiment discussed in the text, with $f_{sky}=2.4\%$. The
uncertainties from the marginalized likelihoods are $\tau=0.12\pm
0.038$ and $A=0.81\pm 0.15 $, given a prior that $\tau>0.04$. The
joint confidence intervals correspond to $1/\sqrt{e}$ (inner) and
$1/e$ (outer) of the maximum likelihood. This experiment can
determine the redshift of the onset of reionization with a
precision of $\delta z \leq 3$ if reionization occurred at
$z>10$.} \label{fig:toytri}
\end{figure}

COSMOMC is a Monte Carlo approach \cite{Lewis02} useful for
evaluating sensitivity to cosmological parameters. Figure
\ref{fig:toytri} shows the parameter estimation results of COSMOMC
for the toy experiment. Due to the cut-sky and the intrinsically
low signal-to-noise ratio of realistic reionization probes, such
as WMAP, it is unlikely that the marginalized likelihood functions
for $\tau$ will be convincingly gaussian in nature. Indeed,
Spergel et al. explicitly points out \cite{spergel2003} that
$\tau$ is the only parameter with a highly non-gaussian form.
COSMOMC facilitates the exploration of the details of the
likelihood curves and the effects of adding priors on the Thomson
optical depth.

\begin{figure}[htb]
\centerline{\psfig{file=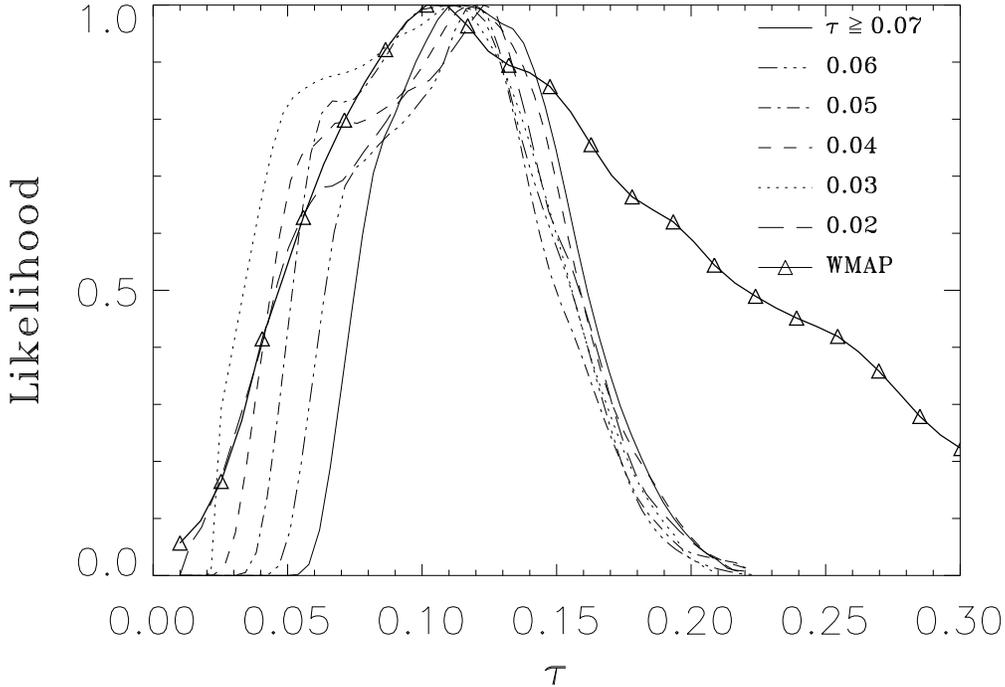, width=5.9in, angle=0}}
\caption{The effects of including prior information on $\tau$ on
the likelihood functions for the toy experiment described in the
text. The likelihoods were computed with simulated data for the
toy experiment, using priors ranging from $\tau>0.02$ to
$\tau>0.07$. The likelihood curve for WMAP with no priors is shown
for comparison as the dashed curve. The upper $1\sigma$ range of
the likelihood curve is robust to changing the prior, whereas the
lower range is more sensitive. } \label{fig:priors}
\end{figure}

Figure \ref{fig:priors} shows the likelihood functions for $\tau$
for the toy experiment with $\Theta=18\arcdeg$ as a function of
$\tau$-priors ranging from $0<\tau<1$ to $0.07 < \tau < 1$ (weaker
constraint than the SDSS Gunn-Peterson observations provide). The
toy-experiment is also compared to the WMAP-only likelihood result
in the same figure.

As discussed above, Kaplinghat \textit{et al.} demonstrate that
the ability of a CMB polarimeter to constrain the redshift of
reionization is $\sigma_{z_{ri}}=333\sigma_\tau/\sqrt{z_{ri}}$.
Thus, conservatively, if reionization is consistent with the WMAP
$1\sigma$ \textit{lower limit}  ($z_{ri}\simeq 13$), the toy
experiment with ($\Theta=18\arcdeg$) could establish (with $>95\%$
confidence) that reionization was earlier than $z=6.28$. This is
the redshift of the SDSS quasar neutral hydrogen fraction
determination from Gunn-Peterson QSO absorption features. We note
that these results are for analysis of the grad-mode data only.
The toy experiment can either use its own data to measure the CMB
temperature anisotropy or WMAP's. Including the
temperature-polarization cross-correlation data the precision on
$\tau$ improves by $\sim 20\%$. Finally, since $\tau$ and the
primordial power spectrum amplitude $A$ are exactly degenerate,
$\sigma_{A}=2\sigma_\tau$, and the toy experiment can achieve
$~6\%$ fractional uncertainty on $A$.

\section{Prospects for the future}

WMAP's on-going measurements of CMB polarization are expected to
continue until at least 2006. In 2007 Planck will be launched and
will start making full-sky observations of the polarization of the
CMB. Planck is expected to achieve near-cosmic variance limit on
$\tau$. NASA's Beyond Einstein initiative features a CMB
polarimeter called the Inflation Probe\cite{nasa}. This instrument
is designed to detect the signature of inflationary gravitational
waves over a wide range of inflationary energy scales. This
experiment will achieve cosmic variance limited precision on
$\tau$, which is essentially the same as Planck's sensitivity
\cite{kaplinghat2003b}. The European Space Agency plans a large
angular scale polarimeter deployed on the International Space
Station called SPORT, which would also be sensitive to
reionization\cite{colombo2004}.

\section{Discussion}
This paper has emphasized that there is more to CMB polarization
than simply the lifting of cosmic degeneracies. Upcoming
experiments will be able to probe the details of reionization and
the signature of the gravitational wave background simultaneously.
We have emphasized that there is far more information in CMB
polarization data than whether or not two power spectra can be
distinguished from one another. However, conservative foreground
mitigation techniques appear to degrade the precision with which
the epoch of reionization, and even the Thomson optical depth
itself, can be obtained by a cosmic variance limited experiment.
Whether such conservatism is warranted is debateable. But when
probing signals below $100n$K level it is clear that foreground
mitigation is vital. More sophisticated techniques may restore
limits closer to the full-sky cosmic variance limit. Detailed
simulations are warranted. A current generation ground based
polarimeter is capable of probing the details of the reionization
epoch as well as the signature of inflationary gravitational
waves. Clearly, high-fidelity CMB polarization observations are
poised to enhance and extend our knowledge of the cosmic Dark
Ages.

\section{Acknowledgements}
CMBFAST, CAMB, and COSMOMC were used in the preparation of this
manuscript. Use of the San Diego Supercomputing Center is
gratefully acknowledged.

\bibliography{Polarization}
\bibliographystyle{unsrt}

\end{document}